\documentclass[prd,12pt,onecolumn,aps,superscriptaddress,nofootinbib,showpacs]{revtex4-1}
\usepackage{amsmath}
\usepackage{amssymb}
\usepackage{graphicx}
\usepackage{color}
\usepackage{epsfig}
\usepackage{dcolumn}
\usepackage{float}
\usepackage{lipsum}
\usepackage{soul}
\usepackage{bm}
\usepackage{times}
\usepackage{multirow}
\usepackage{hyperref}
\hypersetup{
  colorlinks=true,
  citecolor=blue,
  linkcolor=blue,
  urlcolor=blue}

\usepackage{epsfig}
\usepackage{dcolumn}
\usepackage{morefloats}

\newcommand{\ba}{\begin{array}}
\newcommand{\ea}{\end{array}}
\newcommand{\be}{\begin{equation}}
\newcommand{\ee}{\end{equation}}
\newcommand{\beqa}{\begin{eqnarray}}
\newcommand{\eeqa}{\end{eqnarray}}
\def\321{$SU(3)\times SU(2)\times U(1)$}

\hypersetup{
  colorlinks=true,
  citecolor=blue,
  linkcolor=blue,
  urlcolor=blue}
 
\usepackage{epsfig}
\usepackage{dcolumn}

\newcommand{\no}{\nonumber\\}

\begin{document}


\title{\boldmath Consequences of $\mu$-$\tau$ reflection symmetry for $3+1$ neutrino mixing}

\author{Kaustav Chakraborty}
\email[Email Address: ]{kaustav@prl.res.in}
\affiliation{Theoretical Physics Division, Physical Research Laboratory, Ahmedabad - 380009, India}
\affiliation{Discipline of Physics, Indian Institute of Technology, Gandhinagar - 382355, India}

\author{Srubabati Goswami}
\email[Email Address: ]{sruba@prl.res.in}
\affiliation{Theoretical Physics Division, 
Physical Research Laboratory, Ahmedabad - 380009, India}

\author{Biswajit Karmakar}
\email[Email Address: ]{biswajit@prl.res.in}
\affiliation{Theoretical Physics Division, 
Physical Research Laboratory, Ahmedabad - 380009, India}

\begin{abstract}
{We investigate the consequences of $\mu-\tau$ reflection symmetry in presence of a light sterile neutrino for the $3+1$ neutrino mixing  scheme. We discuss the implications of total $\mu-\tau$ reflection symmetry as well partial $\mu-\tau$ reflection symmetry. For the total $\mu-\tau$ reflection symmetry we find that values of $\theta_{23}$ and $\delta$ remains confined near $\pi/4$ and $\pm \pi/2$ respectively. The current allowed region for $\theta_{23}$ and $\delta$ in case of inverted hierarchy lies outside the 
area preferred by the total $\mu-\tau$ reflection symmetry. However, interesting  predictions on the neutrino mixing  angles and Dirac CP violating phases are obtained considering partial $\mu-\tau$ reflection symmetry. We obtain predictive correlations between the neutrino mixing angle $\theta_{23}$ and  Dirac CP phase $\delta$ and study the  testability of these correlations at the future  long baseline experiment DUNE. We find that while the imposition of  $\mu-\tau$ reflection symmetry in the 
first column admits both normal and inverted neutrino mass hierarchy, demanding $\mu-\tau$ reflection symmetry for the second column excludes the inverted hierarchy. Interestingly, the sterile mixing  angle $\theta_{34}$ gets tightly constrained considering the $\mu-\tau$ reflection symmetry in the fourth column. We also study the implications  of $\mu-\tau$ reflection symmetry for the Majorana phases and neutrinoless double beta decay in the 3+1 scenario.
}
\end{abstract}

\maketitle
\section{Introduction}
Over the past years non-zero neutrino masses and mixings have been well established by several neutrino oscillation 
experiments and most of the parameters have been measured with considerable precision. 
The parameters governing the three generation neutrino oscillation phenomena are the three mixing angles 
 (namely, solar mixing angle $\theta_{12}$, atmospheric mixing
angle $\theta_{23}$ and rector mixing angle $\theta_{13}$), two mass-squared differences (namely, solar mass-squared 
difference $\Delta m^2_{sol} = m^2_2 -  m^2_1$ and atmospheric mass-squared difference $\Delta m^2_{atm} = m^2_3 -  m^2_1$)
and Dirac CP phase $\delta$. Among these, the unknown parameters at the present epoch are (a) the octant
of $\theta_{23}$, $i.e.$ $\theta_{23}<45^{\circ}$ (Lower Octant, LO) or $\theta_{23}>45^{\circ}$ (Higher Octant, HO), (b) 
sign of $\Delta m_{atm}^2$, $i.e.$ mass ordering of
neutrinos  where $\Delta m_{atm}^2 > 0$ is called Normal Hierarchy (NH), $\Delta m_{atm}^2 < 0$ is called Inverted Hierarchy (IH) and
(c) magnitude of Dirac CP phase $\delta$. Oscillation experiments are sensitive to the mass-squared differences but the absolute 
mass scale of the light neutrinos are still unknown and there exists only an upper bound
on the sum of absolute neutrino masses $\sum_{i=1}^{3} m_i \leq 0.17$ eV~\cite{Ade:2015xua}, from cosmology.
From the theoretical perspective, lots of effort have been exercised in last
few  decades to realize the observed  neutrino mixing pattern. In this regard, many discrete flavor symmetry groups were
exploited to  understand the dynamics of this mixing pattern in the lepton sector by extending the Standard Model gauge
group with some additional symmetry. A review on lepton masses and mixing 
based on such discrete groups can be found for instance in~\cite{King:2015aea,Altarelli:2010gt,Smirnov:2011jv,Ishimori:2010au, King:2013eh}.

The observational data guided by $\theta_{23}\approx 45^{\circ}$ is indicative of a simple $\mu$-$\tau$ flavor symmetry.  
The simplest realization of such $\mu$-$\tau$ flavor symmetry is 
known as $\mu$-$\tau$ permutation symmetry. Conventionally, $\mu$-$\tau$ permutation symmetry is identified with the  transformation given by
$\nu_{e} \rightarrow \nu_{e}$, $\nu_{\mu} \rightarrow \nu_{\tau}$ and $\nu_{\tau}\rightarrow \nu_{\mu}$, imposition of which leaves the neutrino mass term unaltered. There exists a plethora of models based on 
various discrete flavor symmetry groups possessing an   underlying  $\mu$-$\tau$ permutation symmetry. For example,
with $\sin^2\theta_{23}=1/2, ~\sin^2\theta_{12}=1/3$ and  $\sin^2\theta_{13}=0$ one can obtain a special mixing
pattern known as tribimaximal mixing~\cite{Harrison:2002er}\footnote{Here it's worth  mentioning that mixing schemes 
like trimaximal, bimaximal, golden ratio also depends upon similar hypothesis of the lepton mixing matrix.}.  
Such first approximations of the  neutrino data can easily be reproduced with discrete flavor group like $A_4, S_4$ 
etc\cite{Altarelli:2010gt,King:2011zj,Karmakar:2014dva,Karmakar:2016cvb,Borah:2018nvu}. For a review on $\mu$-$\tau$ flavor symmetry and its phenomenological implications see~\cite{Xing:2015fdg}.

Present oscillation data, particularly  after precise measurement of nonzero 
$\theta_{13}$ ($\sim 8^{\circ}$-$9^{\circ}$), however rules  out exact $\mu$-$\tau$ permutation symmetry and  motivates one to go beyond this symmetry. In this context, a particular variant of  $\mu$-$\tau$ flavor symmetry,  known as $\mu$-$\tau$ reflection symmetry, which predicts both nonzero  $\theta_{13}$ as well as
maximal CP violation as hinted by current observation  is worth studying.
This idea, based on the cumulative operation of $\mu$-$\tau$  flavor exchange and CP transformation was first coined by  Harrison and  Scott~\cite{Harrison:2002et}.
This can be expressed  as the
transformation :  $\nu_{e} \rightarrow \nu_{e}^c$, $\nu_{\mu} \rightarrow \nu_{\tau}^c$ and  $\nu_{\tau} \rightarrow 
\nu_{\mu}^c$ (`c' stands for the charge conjugation of the corresponding neutrino field), under which the neutrino mass
term remains unchanged.  As $\mu$-$\tau$  reflection  symmetry is still phenomenologically viable,  model building with
such underlying symmetry in neutrino sector became popular in recent times, particularly with three active 
neutrinos~\cite{Altarelli:2010gt, Altarelli:2012ss,Smirnov:2011jv,Ishimori:2010au, King:2013eh}. 
The predictions of the $\mu$-$\tau$  reflection symmetry for three neutrino mixing can be summarized as follows:
A) $\theta_{23}= 45^\circ$, $\theta_{13}= 0^\circ$  or B) $\theta_{23}= 45^\circ$, $\delta= 90^\circ ~{\rm or}~ 270^{\circ}$.
Case A is disfavored after measurement on non-zero $\theta_{13}$ by reactor experiments~\cite{An:2012eh}. On the other hand, case B is disfavored by the 
current  data which points towards non-maximal $\theta_{23}$. 
The consequence of such a symmetry for three neutrino mixing scheme have been discussed in several 
occasions~\cite{Grimus:2003yn,Ferreira:2012ri, Grimus:2012hu,  Mohapatra:2012tb, Ma:2015gka, Joshipura:2015dsa, 
Joshipura:2015zla, Joshipura:2016hvn, Nishi:2016wki, Zhao:2017yvw,Liu:2017frs,Xing:2017mkx,Xing:2017cwb,Joshipura:2018rit,
Nath:2018hjx,Zhao:2018vxy}. In particular, breaking of $\mu$-$\tau$ reflection symmetry to generate the deviation from maximal 
$\theta_{23}$ have been considered in \cite{Zhao:2017yvw,Liu:2017frs,Xing:2017mkx,Xing:2017cwb,Joshipura:2018rit,
Nath:2018hjx,Zhao:2018vxy}. 
Another theoretically motivated~\cite{Xing:2014zka, Chakraborty:2018dew,Joshipura:2016quv,Ge:2011ih,Ge:2011qn} scenario  called partial $\mu$-$\tau$ 
reflection symmetry have also been studied to generate deviations from the above values and which resulted in interesting 
correlations \cite{Albright:2008rp,Grimus:2008tt} between mixing parameters. 
All the  discrete
subgroups of $SU(3)$ belonging to class C or D and having three 
dimensional irreducible representation  can lead to the realisation 
of partial $\mu-\tau$ reflection symmetry~\cite{Joshipura:2016quv}.  
Discrete subgroups of $U(3)$ can also serve the same purpose, see~\cite{Joshipura:2016quv,Chakraborty:2018dew} for discussion.
%
%
%

In addition to three active neutrinos, there may exist a light sterile neutrino (Standard Model gauge singlets) at the eV 
scale
(for a review see \cite{Abazajian:2012ys}) 
which can address anomalies in $\bar{\nu_\mu \rightarrow \nu_e}$ oscillations observed in some short-baseline neutrino oscillation experiments. 
Initially the anomaly was found in the 
antineutrino flux measurement of LSND accelerator experiment~\cite{Athanassopoulos:1996jb, Aguilar:2001ty} at Los Alamos which
was 
subsequently confirmed by MiniBooNE~\cite{Aguilar-Arevalo:2013pmq} (a short baseline experiment at Fermilab). 
Very recently MiniBooNE experiment again 
refurbished their earlier results with $\nu_e$ appearance data reinstating the presence of a light sterile neutrino~\cite{Aguilar-Arevalo:2018gpe}.
Results from few experiments like 
gallium solar experiments~\cite{Acero:2007su,  Kopp:2013vaa, Giunti:2012tn} with artificial neutrino sources, 
reactor neutrino experiments\cite{Mention:2011rk, Mueller:2011nm} with recalculated fluxes 
also support the hypothesis of at-least one sterile neutrino.  In this context the 3+1 scenario \cite{Goswami:1995yq} consisting of three active neutrinos and mixing with one eV scale sterile neutrino 
is considered to be most viable \cite{Dentler:2018sju,Giunti:2019aiy}. Here, we have to keep in mind that inclusion of sterile neutrinos must face tight cosmological hurdles coming from the Cosmic Microwave Background 
observations, Big Bang Nucleosynthesis and Large Scale Structures. 
Although fully thermalized sterile neutrinos with mass $\sim$ 1 eV are not cosmologically safe, they can still be generated via `secret interactions'~\cite{Dasgupta:2013zpn,Chu:2018gxk,Song:2018zyl}.
For a brief  review on eV scale sterile neutrinos see~\cite{Abazajian:2017tcc}. Despite many constrains as well as  tension between disappearance and appearance data from oscillation experiments the sterile neutrino conjecture is still a topic of intense research. 

In the context of $3+1$ neutrino mixing exact $\mu$-$\tau$ permutation symmetry would still give $\theta_{13}$ zero. 
Studies have been accomplished in the literature examining the possible role of active-sterile mixing 
in generating a breaking of this symmetry starting  from a $\mu$-$\tau$ symmetric $3\times 3$ neutrino mass matrix \cite{Barry:2011wb,Barry:2011fp,Rivera-Agudelo:2015vza,Merle:2014eja,Xing:2017cwb,Borah:2016fqj,Sarma:2018bgf}.
In this paper we concentrate on the ramifications of  $\mu$-$\tau$ reflection  symmetry for the $4\times 4$ neutrino mass matrix 
in presence  of one sterile neutrino. We study the consequences of total as well as partial $\mu$-$\tau$ reflection  symmetry in 
the 3+1 framework and obtain predictions and correlations between different parameters. We also formulate the $4\times 4$ neutrino
mass matrix which can give rise to such a $\mu$-$\tau$  reflection symmetry. Further we study the experimental consequences of $\mu$-$\tau$ 
reflection symmetry at the future long baseline neutrino oscillations experiment DUNE. In addition we discuss the implications of $\mu$-$\tau$  reflection symmetry for Majorana phases and neutrinoless double $\beta$ decay.

Rest of this paper is organized as follows. In Section \ref{sec1} we first construct the generic structure of the 
$4\times 4$ mass matrix which can give rise to $\mu$-$\tau$  reflection symmetry for sterile neutrinos. In the next section, we find the correlation among the active and sterile mixing angles and Dirac CP phases. In Section \ref{sec4} we study the experimental implications 
of such $\mu$-$\tau$  reflection symmetry for DUNE experiment and also calculate the effective neutrino mass which can be probed through neutrinoless double $\beta$ decay experiments. Then finally in Section \ref{sec5} we summarize the findings.

\section{$\mu$-$\tau$ reflection symmetry for 3+1 neutrino mixing}\label{sec1}
Guided by the atmospheric neutrino data, the $\mu$-$\tau$ reflection symmetry was first proposed for 
3-generation neutrino mixing back in 2002~\cite{Harrison:2002et,Grimus:2003yn}. Under such symmetry the elements of
lepton mixing matrix satisfy :
\begin{equation}\label{umutau}
 |U_{\mu i}|=|U_{\tau i}|~~~~~{\rm where}~~~{i=1,2,3}. 
\end{equation}
This indicates that the moduli of $\mu$ and $\tau$ 
flavor elements of the $3\times 3$ neutrino mixing matrix are equal. 
With these constraints, the neutrino mixing matrix can be parameterised as~\cite{Harrison:2002et,Grimus:2003yn}
\begin{eqnarray}\label{u3b3}
U_{0}&=&\left(
\begin{array}{ccc}
 u_1  & u_2   & u_3 \\
v_1  & v_2   & v_3 \\
v^*_1  & v^*_2   & v^*_3 
\end{array}
\right), 
\end{eqnarray}
where the entries in the first row, $u_i$'s are real (and non-negative)\footnote{Various implications of Majorana phases under such symmetry can be found in \cite{Nishi:2016jqg}.}.
 $v_i$  satisfy the orthogonality condition 
${\rm Re}(v_jv_k^*)=\delta_{jk}-u_k u_k$ \cite{Xing:2015fdg}. 
In~\cite{Grimus:2003yn}, it was argued that the mass matrix leading to  the mixing matrix given in Eq. \ref{u3b3} 
can be written as 
\begin{eqnarray}
\mathcal{M}_0&=&\left( \label{m3b3}
\begin{array}{ccc}
 a  & d   & d^* \\
d  & c   & b \\
d^* & b   & c^*
\end{array}
\right), 
\end{eqnarray}
where $a,b$ are real and $d,c$ are complex parameters. As a consequence of the symmetry given in Eq. \ref{umutau}-\ref{m3b3},
we obtain the  predictions for maximal $\theta_{23}=45^\circ$ and $\delta= 90^\circ ~{\rm or}~ 270^{\circ}$ in the basis where the 
charged leptons are considered to be diagonal. This scheme however still leaves room for nonzero $\theta_{13}$.
Several attempts were made in this direction  to explain correct mixing (for
three active  neutrinos) with $\mu$-$\tau$ reflection symmetry and to study their origin and consequences in various scenarios
\cite{Liu:2017frs, Zhao:2018vxy,Nath:2018xkz,Chakraborty:2018dew,Nath:2018xih,Nishi:2018vlz,Rodejohann:2017lre, Zhao:2017yvw, 
Nishi:2016wki,Chen:2015siy,Joshipura:2015dsa,Zhou:2014sya,He:2015xha,He:2012yt,Ge:2010js,He:2011kn,Nath:2018zoi}. 

Although, $\mu$-$\tau$ reflection symmetry is 
well studied for three active neutrinos, it lacks a comprehensive study considering sterile neutrinos. 
Now such a mixing scheme 
can easily be extended for a $3+1$ scenario incorporating sterile neutrinos. Under such circumstances,
the $4\times 4$ neutrino mixing matrix can be parameterised  as 
\begin{eqnarray}\label{u4b4}
U&=&\left(
\begin{array}{cccc}
u_1  & u_2   & u_3 & u_4\\
v_1  & v_2   & v_3  & v_4\\
v^*_1  & v^*_2   & v^*_3 & v^*_4 \\
w_1  & w_2   & w_3  & w_4\\
\end{array}
\right), 
\end{eqnarray}
where $u_i, w_i$ are real but $v_i$ are complex. Within this extended scenario, the mass matrix can now be written as 
\begin{eqnarray}
\mathcal{M} &=&\left( \label{m4b4}
\begin{array}{cccc}
a   & d   & d^* & e\\
d   & c   & b   & f\\
d^* & b   & c^* & f^*\\
e   & f   & f^* & g  
\end{array}
\right), 
\end{eqnarray}
where $a,b,e,g$ are real and $d,c,f$ are complex parameters. 
Such a complex symmetric mass matrix can be obtained from the Lagrangian 
\be
\mathcal{L}_{\rm mass}
= \frac{1}{2}\, \nu_L^T C^{-1} \mathcal{M}_\nu \nu_L + {\rm H.C.}
\ee
%
with $U^T \mathcal{M}_\nu U = \hat m \equiv \mathrm{diag}
\left( m_1, m_2, m_3, m_4 \right)$, where $m_j$'s are the real positive 
mass eigenvalues. 
Here the matrix $\mathcal{M}$ is characterized by the transformation 
\be\label{mat:S}
S \mathcal{M}_\nu S = \mathcal{M}_\nu^* \quad \mathrm{with} \quad
S = \left( \begin{array}{cccc} 1 & 0 & 0 & 0 \\ 0 & 0 & 1 & 0 \\ 0 & 1 & 0 & 0 \\ 0 & 0 & 0 & 1
\end{array} \right) ,.
\ee
and respects the mixing matrix given in Eq. \ref{u4b4}. To verify this compatibility between the neutrino mixing and mass matrix let us first write mixing matrix as $U = ( c_1, c_2, c_3, c_4 )$ with column vectors $c_j$. Then using the diagonalization relation $U^T \mathcal{M}_\nu U = \mathrm{diag}
\left( m_1, m_2, m_3, m_4 \right)$ one can write
\be
\mathcal{M}_\nu c_j = m_j c_j^\ast\, .
\label{eq:M_vect}
\ee
 Now, using Eq. \ref{mat:S}, we find 
 \be
\mathcal{M}_\nu \left( S c_j^\ast \right) = m_j \left( S c_j^\ast \right)^\ast.
\ee
Following the  above equation, one can therefore find another diagonalizing matrix, $U'=SU^*$. Now it can be shown that if both $U$ and $U'$ satisfy the diagonalization relation $U^T \mathcal{M}_\nu U = \mathrm{diag}  \left( m_1, m_2, m_3, m_4 \right)$ with non-degenerate
mass eigenvalues, then there exists a diagonal unitary matrix $X$ such that
\be
\label{Vcond}
S U^\ast = U X\,,
\ee
 here $X_{jj}$ is an arbitrary phase factor  for $m_j=0$ and $X=\pm1$ for $m_j\neq0$.
Therefore the constraint obtained in Eq. \ref{Vcond} leads 
to \footnote{Following the same approach for 3$\nu$ in \cite{Grimus:2003yn}. }
\begin{equation}\label{umutaus}
 |U_{\mu i}|=|U_{\tau i}|~~~~~{\rm where}~~~{i=1,2,3, 4}. 
\end{equation}
The above equation can also be verified in an alternate way. Let us first define an Hermitian matrix as, 
\be
H = \mathcal{M}_\nu^\ast \mathcal{M}_\nu 
\label{eq:herm-mat}
\ee
considering the form of $\mathcal{M}_\nu$ given in Eq. \ref{m4b4} one can easily find
\be
H_{\mu \mu} = H_{\tau \tau} \quad {\rm and} \quad
H_{e \mu} = H_{e \tau}^\ast, \quad
H_{s \mu} = H_{s \tau}^\ast\, .
\label{eq:herm-mat-reln-mutau}
\ee
Now, one can write the diagonalization relation in this case as : $
 H_{\alpha \beta}  = U_{\alpha i} \hat{m}^2_{ij} U_{j \beta}^\dagger$. Hence using Eq (\ref{eq:herm-mat-reln-mutau}) we get 
\be \no
 \sum_{i=1}^{4} \hat{m}^2_{ii} |U_{\mu i}|^2  = \sum_{i=1}^{4} \hat{m}^2_{ii} |U_{\tau i}|^2 \\
\ee
which follows only if masses are degenerate or $|U_{\mu i}| = |U_{\tau i}|$~\cite{Grimus:2003yn}. Therefore, it is now clear to us that 
the  mass matrix given in Eq. \ref{m4b4} actually leads to a mixing matrix of the form in Eq. \ref{u4b4}. In the following section we discuss the consequences of this $\mu$-$\tau$ reflection symmetry involving the active and sterile mixing angles and phases in details.

 It is important to note that the mixing matrix given in Eq. \ref{u3b3} should correspond to the standard neutrino mixing 
 matrix $U_{PMNS}$ for three generation  case. Now, depending upon the choice of the arbitrary phase factor $X$ given in Eq. \ref{Vcond} the Majorana 
 phases can be fixed in the context of $\mu$-$\tau$ reflection symmetry. With the choice of $X_{ii}=$ 1 or -1 the Majorana phases are fixed at 0$^\circ$ or 
 $90^{\circ}$~\cite{Xing:2015fdg, Zhao:2017yvw}. Such fixed values of phases can have  implication for  neutrinoless double beta decay which will be discussed later.


\section{\boldmath Constraining 3+1 neutrino mixing with $\mu$-$\tau$ reflection symmetry}\label{sec2}

For 3+1 neutrino mixing scenario the neutrino mixing matrix $U$ can be written in terms of a  $4\times 4$ unitary matrix. This unitary matrix can be parameterized by three active neutrino mixing angles $\theta_{13}, \theta_{12}, \theta_{23}$ and three more angles originating from active-sterile mixing, namely, $\theta_{14}, \theta_{24}$ and $ \theta_{34}$. It will also contain three Dirac CP violating phases, such as,   $\delta, \delta_{14}$ and $ \delta_{24}$. Hence this  $4\times 4$ unitary PMNS matrix $U$ can be given by 
\be \label{udef}
 U=R_{34}\tilde{R}_{24}\tilde{R}_{14}{R}_{23}\tilde{R}_{13}R_{12}, 
\ee
where the rotation matrices $R$ and $\tilde{R}$'s can be written as 
\begin{eqnarray}\label{Rs}
R_{34}=\left(
\begin{array}{cccc}
1  & 0   & 0 & 0\\
0  & 1   & 0 & 0\\
0  & 0   & c_{34} & s_{34} \\
0  & 0   & -s_{34}  & c_{34}\\
\end{array}
\right), ~~\tilde{R}_{24}=\left(
\begin{array}{cccc}
1  & 0   & 0  & 0\\
0  & c_{24}   & 0  & s_{24}e^{-i\delta_{24}}\\
0  & 0   & 1 &0 \\
0  & -s_{24}e^{i\delta_{24}}   & 0  & c_{24}\\
\end{array}
\right), \nonumber\\
\tilde{R}_{14}=\left(
\begin{array}{cccc}
c_{14}  & 0   & 0 & s_{14}e^{-i\delta_{14}}\\
0  & 1   & 0  & 0\\
0  & 0   & 1 &0 \\
-s_{14}e^{-i\delta_{14}}  & 0   & 0  & c_{14}\\
\end{array}
\right),~~R_{23}=\left(
\begin{array}{cccc}
1  & 0   & 0 & 0\\
0  & c_{23}   & s_{23} & 0\\
0  & -s_{23}   & c_{23} & 0 \\
0  & 0   & 0  & 1\\
\end{array}
\right), \nonumber\\
\tilde{R}_{13}=\left(
\begin{array}{cccc}
c_{13}  & 0   & s_{13}e^{-i\delta} & 0\\
0  & 1   & 0  & 0\\
-s_{13}e^{i\delta}  & 0   & c_{13} &0 \\
0  & 0   & 0  & 1\\
\end{array}
\right),~~R_{12}=\left(
\begin{array}{cccc}
c_{12}  & s_{12}   & 0 & 0\\
-s_{12}  & c_{12}   & 0 & 0\\
0  & 0   & 1 & 0 \\
0  & 0   & 0  & 1\\
\end{array}
\right).
\end{eqnarray}
Along with the parameterization  defined in Eq. \ref{udef}, there also exists a diagonal phase matrix,
$P = \rm{diag}(1, e ^{i\alpha} , e^{ i(\beta+\delta )} , e ^{i(\gamma+\delta_{14} )})$, where $\alpha, \beta$ and $\gamma$ 
are the Majorana phases. The PMNS matrix with the Majorana phases takes the form as
\be \label{maj-U4}
 U=R_{34}\tilde{R}_{24}\tilde{R}_{14}{R}_{23}\tilde{R}_{13}R_{12} P, 
\ee
 Note that, the correspondence of the  mixing matrix in Eq. \ref{u4b4} along with the diagonal phase matrix $P$ in Eq. \ref{maj-U4} implies that the Majorana phases are zero or $\pm \frac{\pi}{2}$.
However, in light of Eq. \ref{umutau} this diagonal phase matrix do not play any role in the present analysis. But they can play role in neutrinoless double $\beta$ decay which will be discussed in Section \ref{sec4}.
Following these conditions, one can obtain four different equalities among six mixing angles and three Dirac CP violating phases. To keep the present analysis simple, first  we have assumed the sterile Dirac CP violating phases ($\delta_{14}$ and $ \delta_{24}$) to be zero. For this case, $\delta_{14}$ = $ \delta_{24}$ = $0^\circ$,  from Eq. \ref{umutau} these four correlations can be written as, 
\begin{eqnarray}
 \cos\delta&=&\frac{
                  (a_1^2+b_1^2)-(c_1^2+d_1^2)
                  }
                  {
                  2(c_1 d_1-a_1 b_1)
                  }\label{eq:11}~~(\rm{for}~|U_{\mu 1}|=|U_{\tau 1}|)\\
\cos\delta&=&\frac{
                  (a_2^2+b_2^2)-(c_2^2+d_2^2)
                  }
                  {
                  2(a_2 b_2-c_2 d_2)
                  }\label{eq:22}~~(\rm{for}~|U_{\mu 2}|=|U_{\tau 2}|)        \\
\cos\delta&=&\frac{
                  (a_3^2+b_3^2)-(c_3^2+d_3^2)
                  }
                  {
                  2(a_3 b_3-c_3 d_3)
                  }\label{eq:33} ~~(\rm{for}~|U_{\mu 3}|=|U_{\tau 3}|) \\
 \tan^2\theta_{24}&=&\sin^2\theta_{34}\label{eq:44}    \hspace{2.5cm}~~(\rm{for}~|U_{\mu 4}|=|U_{\tau 4}|)               
\end{eqnarray}
where 
\begin{eqnarray}
 a_1&=&(c_{12}s_{13}s_{23}s_{24}s_{34}-c_{12}c_{23}c_{34}s_{13}),
 b_1=(c_{34}s_{12}s_{23}-c_{12}c_{13}c_{24}s_{14}s_{34}+c_{23}s_{12}s_{24}s_{34})\nonumber\\
 c_1&=&(c_{23}c_{24}s_{12}+c_{12}c_{13}s_{14}s_{24}),
 d_1=(c_{12}c_{24}s_{13}s_{23}),
  a_2=c_{12}(c_{34}s_{23}+c_{23}s_{24}s_{34})+s_{12}c_{13}c_{24}s_{14}s_{34}\nonumber\\
 b_2&=&s_{12}(s_{13}s_{23}s_{24}s_{34}-c_{23}c_{34}s_{13}),
 c_2=(c_{12}c_{23}c_{24}-s_{12}c_{13}s_{14}s_{24}),
 d_2=s_{12}c_{24}s_{13}s_{23}\nonumber\\
 a_3&=&c_{13}(c_{23}c_{34}-s_{23}s_{24}s_{34}),
 b_3=c_{24}s_{13}s_{14}s_{34}),
 c_3=s_{13}s_{14}s_{24}~ {\rm and }~
 d_3=c_{13}c_{24}s_{23}.\label{corpara}
\end{eqnarray}
Here, the first three equalities enable us to study the correlation among  the mixing angles $\theta_{12}, \theta_{23}, \theta_{13}, \theta_{14}, \theta_{24}, \theta_{34}$  and Dirac CP phase $\delta$ whereas the fourth relation yields a crucial correlation between the two sterile mixing angles $\theta_{24}$ and  $\theta_{34}$. For $\delta_{14}~ \&~ \delta_{24} \neq 0^\circ$, such compact expressions cannot be obtained. However, in our numerical analysis we have studied the effect of inclusion of these phases. When the sterile mixing angles ($\theta_{14}, \theta_{24}, \theta_{34}$) are taken to be zero, the correlations obtained in Eq. \ref{eq:11}-\ref{corpara} reduces to the three neutrino mixing scenarios studied in \cite{Chakraborty:2018dew,Petcov:2018snn,Albright:2008rp,Grimus:2008tt,Girardi:2016zwz} .
\begin{table}[h!]
\begin{center}
\begin{tabular}{|c|c|c|}
\hline
Oscillation parameters & Best-fit & $3 \sigma$ range \\
\hline \hline
$\theta_{13}$ & $8.6^\circ$  &  $8.2^\circ : 9.0^\circ$ \\
\hline
$\theta_{12}$ & $33.8^\circ$  &  $31.6^\circ : 36.3^\circ$ \\
\hline
$\theta_{23}$ & $49.5$  & $40^\circ : 52^\circ$ \\
\hline
$\Delta m^2_{21}$ (eV$^2$) & $7.4\times 10^{-5}$ & fixed \\
\hline 
$|\Delta m^2_{31}|$ (eV$^2$) & $2.5\times 10^{-3}$ & $(2.35 : 2.65)\times 10^{-3}$  \\
\hline
$\delta$  & $0^\circ : 360^\circ$ & $0^\circ : 360^\circ$ \\
\hline \hline
Oscillation parameters & Representative Value & $3 \sigma$ range \\
\hline \hline
$\Delta m^2_{41}$ (eV$^2$) & $1$ & fixed  \\
\hline
$\theta_{14}$ & $9^\circ$   & $4^\circ : 10^\circ$ \\
\hline
$\theta_{24}$ & $9^\circ$  & $5^\circ : 10^\circ$ \\
\hline
$\theta_{34}$ & $9^\circ$ & $0^\circ : 11^\circ$ \\
\hline
$\delta_{\rm{14}}$  & -- & $0^\circ : 360^\circ$ \\
\hline
$\delta_{\rm{24}}$  & -- & $0^\circ : 360^\circ$ \\
\hline
\end{tabular}
\caption{The best-fit values and 3$\sigma$ ranges of the 3 neutrino oscillation parameters \cite{nufit,Esteban:2018azc} used in the present analysis and the representative ranges for 3+1 neutrino mixing \cite{Gariazzo:2017fdh}. }
\label{tab:data1}  
\end{center}
\end{table}
Below we study the various correlations between the parameters due to Eqs.\ref{eq:11}-\ref{eq:44}. 
We find that the phenomenologically interesting correlations are between $\theta_{23}-\delta$ and $\theta_{24}-\theta_{34}$. 
Since the other sterile mixing angles are already restricted to a narrow range, no other important correlations are obtained.

\subsection{\boldmath Total $\mu-\tau$ symmetry}
\begin{figure}[h!]
$$
\includegraphics[height=6.5cm]{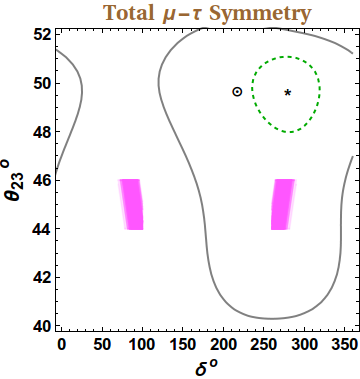}
$$
\caption{Allowed region for total $\mu-\tau$ symmetry for 3+1 neutrino scenario. Here the other mixing parameters are varied within $3\sigma$ range~\cite{nufit,Esteban:2018azc,Gariazzo:2017fdh}}
\label{fig:comp}
\end{figure}

In this subsection we present the results assuming $\mu - \tau$ reflection symmetry to be valid for all the four columns simultaneously; we call this as the total $\mu-\tau$ reflection symmetry. 
The magenta shaded region in Fig. \ref{fig:comp} represents the allowed region for total  $\mu-\tau$ symmetry for 3+1 neutrino scenario in $\theta_{23}-\delta$ plane. The analysis is performed by varying the other mixing parameters in their $3\sigma$ range as in Tab. \ref{tab:data1} and the sterile CP phases $\delta_{14}$ and $\delta_{24}$ between $0^{\circ}$ to $360^{\circ}$\footnote{Here (and in the rest of the analysis, unless otherwise mentioned) we vary both $\delta_{14}$ and $\delta_{24}$ between $0^{\circ}$ to $360^{\circ}$.}. 
The grey solid and green dashed contours denote the currently allowed parameter space for NH and IH respectively in this and the subsequent figures.
The application of the total  $\mu-\tau$ symmetry significantly restricts the parameters $\theta_{23}$ and $\delta$. $\theta_{23}$ is primarily restricted around the maximal while  $\delta$ falls in the close vicinity of $90^{\circ}$ and $270^{\circ}$. Comparing the results with 3 neutrino scenario \cite{Xing:2014zka, Chakraborty:2018dew} where $\theta_{23}$ is strictly restricted to be maximal and $\delta$ to $90^{\circ}$ and $270^{\circ}$ we conclude that the involvement of the sterile mixing angles and phases lead to slight deviations in $\theta_{23}$ and $\delta$ from their 3 generation predictions. However, the current global fit results from \cite{nufit} suggests that the best fit for $\theta_{23}$ is $49.5^\circ$ for both normal and inverted hierarchies. 
Thus, even with inclusion of sterile neutrinos, total $\mu - \tau$ reflection symmetry cannot explain the current best-fit. This motivates us to consider the partial $\mu-\tau$ reflection symmetry for the 3+1 scenario.

\subsection{\boldmath Partial $\mu-\tau$ reflection symmetry}
In this section we discuss the implications partial $\mu-\tau$ reflection symmetry which implies that the condition $|U_{\mu i}|=|U_{\tau i}|$ is satisfied for individual columns. 

\subsubsection{$|U_{\mu 1}|=|U_{\tau 1}|$}

\begin{figure}[h!]
$$
\includegraphics[height=6.5cm]{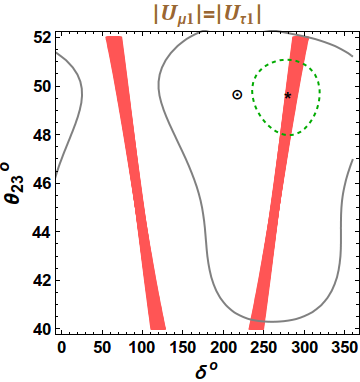}
\hspace*{0.2cm}
\includegraphics[height=6.5cm]{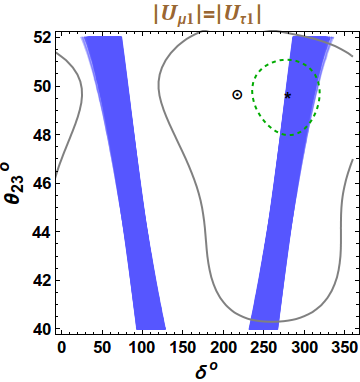}
$$
\caption{Correlation between $\theta_{23}$ and Dirac CP phase $\delta$ for $|U_{\mu 1}|=|U_{\tau 1}|$ with $\delta_{14}$,~$\delta_{24}=0^\circ$ (left panel) and $\delta_{14}$,~$~\delta_{24}\neq 0^\circ$ (right panel) respectively.
Here all other mixing parameters are varied within their $3\sigma$ range as given in Tab. \ref{tab:data1}. The continuous and dashed contours represent $3\sigma$ allowed range in the $\theta_{23}-\delta$ plane and the $\odot$ and $\star$ represent the best-fit values for normal and inverted neutrino mass hierarchy respectively.}
\label{fig:11}
\end{figure}

The correlation obtained from the equality $|U_{\mu 1}|=|U_{\tau 1}|$ has been plotted in the $\theta_{23} - \delta$ plane in Fig. \ref{fig:11}. The left panel with red contours in Fig. \ref{fig:11} represents the case with sterile phases $\delta_{14}$ and $\delta_{24}$ taken to be zero while the  right panel with blue contours denotes the case with $\delta_{14}$ and $\delta_{24}$ to be non-zero.  In these panels, the grey continuous and green dashed contours represent $3\sigma$ allowed range in the $\theta_{23}-\delta$ plane and the $\odot$ and $\star$ represents the best-fit values for normal and inverted neutrino mass hierarchy respectively \footnote{Similar descriptions are also true for the subsequent figures for $|U_{\mu 2}|=|U_{\tau 2}|$ and $|U_{\mu 3}|=|U_{\tau 3}|$. }.
The interesting result obtained from these correlations are the range of $\delta$ allowed by the $\mu-\tau$ reflection symmetry. In this case the CP conserving values of $\delta$ are ruled out and preference is seen for maximal CP violation. This points towards an important consequence of $\mu-\tau$ reflection symmetry i.e. if $\mu-\tau$ reflection symmetry is true for the first column of lepton mixing matrix CP violation is implied. The recent global fit~\cite{nufit} result also point towards the maximal CP violation with with preference for $\delta\sim 270^\circ$. 
From both these panels it is clear that under $\mu-\tau$ reflection symmetry in the first column of lepton mixing matrix, inverted hierarchy of neutrino mass is a more favored scenario given the current best-fit values, this is true with or without the involvement of sterile CP phases ($\delta_{14}$,$\delta_{24}$). The inclusion of the sterile CP phases $\delta_{14}$,$\delta_{24}$ predicts slightly larger allowed range for $\delta$.

\subsubsection{$|U_{\mu 2}|=|U_{\tau 2}|$}

\begin{figure}[h!]
$$
\includegraphics[height=6.5cm]{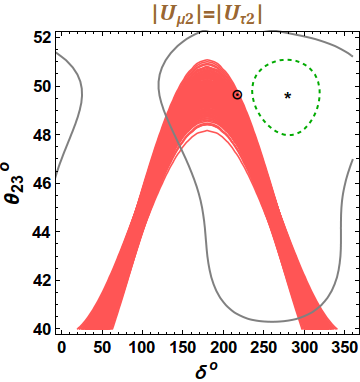}
\hspace*{0.2cm}
\includegraphics[height=6.5cm]{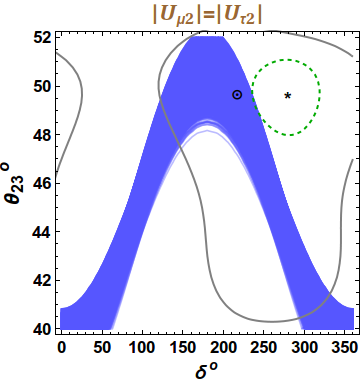}
$$
\caption{Correlation between $\theta_{23}$ and Dirac CP phase $\delta$ for $|U_{\mu 2}|=|U_{\tau 2}|$ with $\delta_{14}$,$\delta_{24}=0^\circ$ (left panel) and $\delta_{14}$, $\delta_{24}\neq 0^\circ$ (right panel) respectively.
Here all other mixing parameters are varied within their $3\sigma$ range as given in Tab. \ref{tab:data1}. The continuous and dashed contours represent $3\sigma$ allowed range in the $\theta_{23}-\delta$ plane and the $\odot$ and $\star$ represents the best-fit values for normal and inverted neutrino mass hierarchy respectively.}
\label{fig:22}
\end{figure}

$\mu-\tau$ reflection symmetry in the second column of lepton mixing matrix (given by $|U_{\mu 2}|=|U_{\tau 2}|$) is plotted in Fig. \ref{fig:22}. It is seen from the left panel of this figure that this symmetry disfavors $\delta = 0^\circ$ while $\delta = 180^\circ$ is allowed from the correlation for the current $\theta_{23}$ range when sterile CP phase are zero. The presence of the sterile CP phases $\delta_{14}$,~$\delta_{24}$ predicts slightly larger allowed range for both $\theta_{23}$ and $\delta$, and in presence of these phases,  both $\delta = 0^\circ$ and $\delta = 180^\circ$ become admissible unlike the previous case in Fig. \ref{fig:11}.
Interestingly the current 3$\sigma$ global fit contours for inverted hierarchy do not overlap with the the allowed region due to $\mu-\tau$ reflection symmetry in the second column of the lepton mixing matrix with sterile CP phases $\delta_{14}$ = $\delta_{24}$ = $0^\circ$. But, there is a slight overlap between the above specified regions once $\delta_{14}$ and $\delta_{24}$ are taken non-zero.

\subsubsection{$|U_{\mu 3}|=|U_{\tau 3}|$}

\begin{figure*}[h!]
$$
\includegraphics[height=6.5cm]{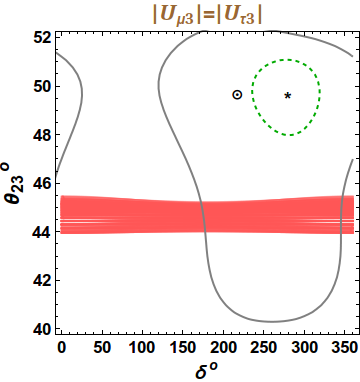}
\hspace*{0.2cm}
\includegraphics[height=6.5cm]{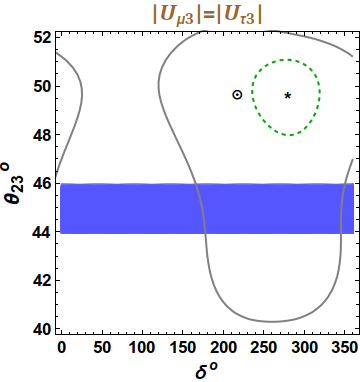}
$$
\caption{Correlation between $\theta_{23}$ and Dirac CP phase $\delta$ for $|U_{\mu 3}|=|U_{\tau 3}|$ with $\delta_{14}$,$\delta_{24}=0^\circ$ (left panel) and $\delta_{14}$, $\delta_{24}\neq 0^\circ$ (right panel) respectively.
Here all other mixing parameters are varied within their $3\sigma$ range as given in Tab. \ref{tab:data1}. The continuous and dashed contours represent $3\sigma$ allowed range in the $\theta_{23}-\delta$ plane and the $\odot$ and $\star$ represents the best-fit values for normal and inverted neutrino mass hierarchy respectively.}
\label{fig:123-2}
\end{figure*}

In Fig. \ref{fig:123-2}  we show the consequence of $|U_{\mu 3}|=|U_{\tau 3}|$  , and it is seen that the allowed value of $\theta_{23}$ stays close to maximal with slight deviation  within a very narrow range 
 with a preference for the lower octant as given in the left panel of Fig. \ref{fig:123-2} with $\delta_{14}$,$\delta_{24}=0^\circ$. 
If we introduce the non-zero values for the sterile CP phases ($\delta_{14}$, $\delta_{24}\neq 0^\circ$), the variation of $\theta_{23}$ remains in the vicinity of $45^\circ$ with equal deviations in both lower and higher octants as evident from the right panel of Fig. \ref{fig:123-2}. However, the deviation is not enough to reach the best-fit $\theta_{23}$ from current data~\cite{nufit,deSalas:2017kay,Capozzi:2018ubv}.
However, if future data from T2K or NO$\nu$A give a value closer 
to maximal $\theta_{23}$ (but not exactly maximal) then this scenario can be
preferred over the three generation case.   
Since for the three flavor case, the condition 
$|U_{\mu 3}|=|U_{\tau 3}|$ leads to $\theta_{23}= 45^\circ$. 
However, in presence of sterile neutrinos there is a  spread 
around the maximal value and future measurements of $\theta_{23}$ 
can confirm or falsify if this condition can indeed be satisfied. 
In this context it is also worthwhile to discuss 
to what extent future high statistics experiments can determine the octant of  $\theta_{23}$ close to maximal value. For instance, it was shown in 
\cite{Ballett:2016daj} from a combined analysis of DUNE and T2HK 
that the octant of $\theta_{23}$ will remain unresolved
for true values in the range $43^\circ - 48.7^\circ$. 
The maximum allowed range of  $\theta_{23}$ for $|U_{\mu 3}|=|U_{\tau 3}|$ in presence 
of sterile mixing and phases being $44^\circ- 46^\circ$, the octant will remain 
undetermined in this situation even with the future high statistics experiments.

\subsubsection{$|U_{\mu 4}|=|U_{\tau 4}|$}
\begin{figure}[h!]
$$
\includegraphics[height=6.5cm]{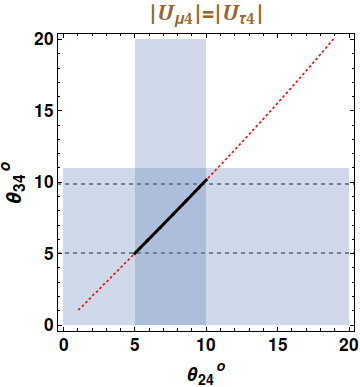}
$$
\caption{Constraints on $\theta_{34}$ obtained from $|U_{\mu 4}|=|U_{\tau 4}|$. Here the red dashed line (including the black line) represents this correlation. Shaded regions are current allowed range~\cite{Gariazzo:2017fdh}. Here 3$\sigma$ allowed range of $\theta_{24}$~\cite{Gariazzo:2017fdh} restricts $\theta_{34}$ within the range $4.9^{\circ}-9.8^{\circ}$ (as given
by the dark black line).}
\label{fig:44s}
\end{figure}

In this 3+1 neutrino framework, the fourth  equality $|U_{\mu 4}|=|U_{\tau 4}|$ establishes a powerful correlation between the two sterile mixing angles
$\theta_{24}$ and $\theta_{34}$. Here we  obtain a one-to-one correspondence between $\theta_{24}$ and $\theta_{34}$ as given in Eq. \ref{eq:44}. Even with the 
involvement of sterile CP phase this relation remains same as evident from Eq. \ref{umutaus} and \ref{udef}.  This correlation yields a linear dependence between
the two sterile mixing angles $\theta_{24}$ and $\theta_{34}$ as given in Fig. \ref{fig:44s}. Once we impose the constraints coming from the the current allowed
value of $\theta_{24}$~\cite{Gariazzo:2017fdh}, the sterile mixing angle $\theta_{34}$ becomes restricted from below significantly. Note that so far there only
exists a upper limit on $\theta_{34}$~\cite{Gariazzo:2017fdh}.
In Fig. \ref{fig:44s} we plot this correlation and find that $\theta_{34}$ lies within the range $4.9^{\circ}-9.8^{\circ}$  corresponding to the 
3$\sigma$ allowed range of $\theta_{24}$~\cite{Gariazzo:2017fdh}. 
Therefore, the $\mu-\tau$ reflection symmetry presented  here restricts the sterile mixing angle $\theta_{34}$ considerably.  
The allowed parameter space in the $\theta_{24} - \theta_{34}$ plane also 
gets restricted. 
 This is one of the most crucial finding in this  $\mu-\tau$ reflection symmetric framework for 3+1 neutrino scenario. 
It is to be noted that among the current constraints on sterile mixing angles, 
the bound on 
$\theta_{34}$ is much weaker, there being only an upper limit on this. 
The reactor neutrino experiments are sensitive to the mixing matrix element 
$U_{e4}^2$ or $s_{14}^2$ in our parametrization.  The short baseline 
oscillation experiments using appearance channel are sensitive to 
the product $|U_{e4}|^2 |U_{\mu 4}^2|$ which contains the product 
$s_{14}^2 s_{24}^2$. Bounds on $\theta_{34}$ have been obtained 
 from atmospheric 
neutrinos at SuperKamiokande \cite{Abe:2014gda}, 
DeepCore detector at Icecube \cite{Aartsen:2017bap},  
and from Neutral Current data
at MINOS \cite{MINOS:2016viw}, NO$\nu$A \cite{ Adamson:2017zcg} and T2K 
\cite{Abe:2019fyx} 
experiments. The  constraints  
on $\theta_{34}$ from individual experiments are somewhat weaker (in the ballpark of $20^\circ - 30^\circ$) 
than what  is obtained in the 
global analysis of \cite{Gariazzo:2017fdh}. In our analysis the later has been
used.  
Neutral current events at the DUNE 
detector can also improve on the bound on $\theta_{34}$ coming 
from a single experiment \cite{Gandhi:2017vzo}.
The $\nu_{\tau}$ appearance channel is also sensitive to $\theta_{34}$ and 
the potential of 
DUNE experiment to constrain this mixing angle has been 
studied in \cite{deGouvea:2019ozk, Ghoshal:2019pab,Coloma:2017ptb}.
Thus it is expected that future data  
can test  this correlation and the allowed parameter space.

These discussions lead us to the inference that partial $\mu-\tau$ reflection symmetry is more favorable scenario. However it is to be noted that in this scenario the case $|U_{\mu 3}|=|U_{\tau 3}|$ is disfavored because it fixes $\theta_{23}$ around the maximal value. Again, the simultaneous application of equalities  $|U_{\mu 1}|=|U_{\tau 1}|$ and $|U_{\mu 2}|=|U_{\tau 2}|$ restricts $\theta_{23} \sim 45^\circ$ hence both these equalities cannot be satisfied together. Such experimental constraints do not apply on $|U_{\mu 4}|=|U_{\tau 4}|$ therefore this equality may still hold. So, the favorable scenarios are:
\begin{itemize}
\item $|U_{\mu 1}|=|U_{\tau 1}|$ with $|U_{\mu 4}|=|U_{\tau 4}|$
\item $|U_{\mu 2}|=|U_{\tau 2}|$ with $|U_{\mu 4}|=|U_{\tau 4}|$
\end{itemize}

\section{Experimental consequences of $\mu$-$\tau$ reflection symmetry}\label{sec4}

\subsection{Neutrino Oscillations Experiments}
In this section we explore the consequences of partial $\mu-\tau$ reflection symmetry in the 3+1 scenario for the Deep Underground Neutrino Experiment (DUNE).

\subsubsection{Experimental and Simulation details}
DUNE is a proposed future long baseline accelerator experiment which is expected to lead the endeavor in determination of the unknown neutrino oscillation parameters. The neutrino source for DUNE is proposed to be the Long Baseline Neutrino Facility (LBNF) at Fermilab, which will provide intense 1.2 MW neutrino beams. The detector is a 40 kt liquid Argon detector located at South Dakota with a baseline of 1300 km. The total POT is expected to be 10 $\times$ $10^{21}$ over a period of 10 years with 5 years each of neutrino and anti-neutrino run. The simulation have been performed using the package General Long Baseline Experiment Simulator (GLoBES) \cite{Huber:2004ka,Huber:2007ji}, and the sterile neutrino effects have been applied using the sterile neutrino engine as described in \cite{Kopp:2007mi}.

To test the correlations at DUNE we define $\chi^2$ as 
\begin{eqnarray}
\chi^2_{{\rm tot}} = \underset{\xi, \omega}{\mathrm{min}} \lbrace \chi^2_{{\rm stat}}(\omega,\xi) + \chi^2_{{\rm pull}}(\xi)  \rbrace.
 \label{chi-tot}
\end{eqnarray}
where, the statistical $\chi^2$ is $\chi^2_{{\rm stat}}$ while the systematic uncertainties are incorporated by $\chi^2_{pull}$. The later is calculated 
by the method of pulls with pull variables given by $\xi$ \cite{pulls_gg, pull_lisi,ushier}. The  oscillation parameters \{$\theta_{23},\theta_{12},\theta_{13}, \delta_{CP}, \Delta m^2_{21},\Delta m^2_{31},\theta_{14},\theta_{24},\theta_{34},\delta_{14}\delta_{24}$\} are represented by $\omega$.
The statistical $ \chi^{2}_{stat} $ is calculated assuming Poisson distribution,
\begin{equation}\label{eq:stat_chisq}
 \chi^{2}_{stat} = \sum_i 2\left( N^{test}_i-N^{true}_i - N^{true}_i \log\dfrac{N^{test}_i}{N^{true}_i}\right).
\end{equation}
Here, `i' stands for the number of bins and $ N^{test}_i, N^{true}_i $ stands for 
total number of test and true events respectively. 
To include the effects of systematics in $ N^{test}_i$, pull and ``tilt" variables are incorporated as follows:
\begin{eqnarray}
N^{{\rm (k)test}}_i(\omega, \xi) = \sum\limits_{k = s, b} N^{(k)}_{i}(\omega)[1 + c_{i}^{(k) norm} \xi^{(k) norm} + c_{i}^{(k) tilt} \xi^{(k) tilt} \frac{E_i - \bar{E}}{E_{max} - E_{min}}] \;,
\end{eqnarray}
where $k = s(b)$ represent the  signal(background) events.
The effect of the pull variable $\xi^{norm}$(${ \xi}^{tilt}$) on the number of events are denoted by $c_i^{norm}$(${c_i}^{tilt}$). The bin by bin mean reconstructed energy is represented by $E_i$ where $i$ represents the bin. $E_{min}$, $E_{max}$ and $\bar{E} = ({E_{max} +E_{min}})/{2}$ are the minimum energy, maximum energy and the mean energy over this range.
 The signal normalization uncertainties used are as follows: 
for $\nu_e$/$\bar{\nu_{e}}$ - 2\% and $\nu_{\mu}$/$\bar{\nu_{\mu}}$ - 5\%. While the background uncertainties vary from 5\% to 20\%.

\subsubsection{$\mu$-$\tau$ reflection symmetry at DUNE for 3+1 neutrino mixing}

\begin{figure}[h!]
 $$
 \includegraphics[height=6.5cm]{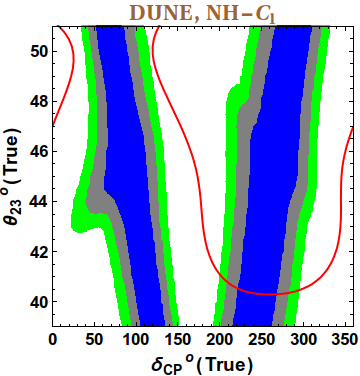}
 \hspace*{0.2cm}
 \includegraphics[height=6.5cm]{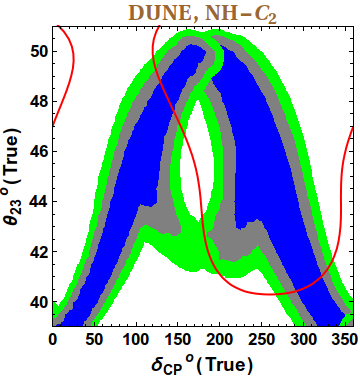}
 $$
\caption{Experimental constraints on $\delta_{CP}$ from the correlations $|U_{\mu 1}|=|U_{\tau 1}|$ and $|U_{\mu 2}|=|U_{\tau 2}|$ for DUNE. The left(right) column indicate correlation $|U_{\mu 1}|=|U_{\tau 1}|$($|U_{\mu 2}|=|U_{\tau 2}|$). The blue, grey and green shaded regions depict 1$\sigma$, 2$\sigma$ and 3$\sigma$ confidence regions respectively. While the red contours represent the currently allowed 3$\sigma$ region considering true Normal Hierarchy.}
\label{fig:exp-verf}
\end{figure}
The consequences of the partial $\mu - \tau$ reflection symmetry at the experiment is observed by analyzing the confidence region in the $\theta_{23}$(true) vs $\delta_{CP}$(true) plane which remains allowed after the application of the symmetry relations in the test parameters.
The approach taken for the numerical analysis can be summarized as follows:
\begin{itemize}
\item The simulated data for the experiment DUNE are generated for the representative true values of the oscillation parameters as given by the best-fit in Tab. \ref{tab:data1} excepting for $\theta_{23}$ and $\delta$. The true values of these parameters are varied over the range $39^\circ  -  51^\circ$ and $0^\circ - 360^\circ$ respectively. The true values of $\delta_{14}$ \& $\delta_{24}$ are taken as $0^\circ$. 

\item The test events are generated by marginalization of the parameters $\theta_{12}$, $\theta_{13}$, $|\Delta m ^2 _{31}|$, $\theta_{23}$, $\theta_{14}$, $\theta_{24}$, $\theta_{34}$ over the range given in Tab. \ref{tab:data1} subject to the condition embodied in Eq. \ref{eq:11} for the left panel and Eq. \ref{eq:22} for the right panel. The other parameters are held fixed at their true values for calculation of $N_{test}$.
In this study we have assumed normal hierarchy as the true hierarchy. We have checked that marginalizing over test hierarchy do not have significant effect because the correlations are independent of hierarchy.

\item For each true value of  $\theta_{23}$ and $\delta$ the $\chi^2$ is minimized and the allowed regions defined by $\chi^2 \le \chi^2_{min} + \Delta \chi^2$ are plotted corresponding to 1$\sigma$, 2$\sigma$ and 3$\sigma$ values of $\Delta \chi^2$.
\end{itemize} 

The experimental consequences at DUNE are presented in Fig. \ref{fig:exp-verf}. Here the first(second) column  represents the correlation $|U_{\mu 1}|=|U_{\tau 1}|$($|U_{\mu 2}|=|U_{\tau 2}|$). The condition $|U_{\mu 4}|=|U_{\tau 4}|$ is also incorporated in both the plots.
Each plot consists of 1$\sigma$, 2$\sigma$ \& 3$\sigma$ confidence regions considering partial $\mu$ - $\tau$ reflection symmetry which are shaded as blue, grey and green respectively. The current 3$\sigma$ permitted region from NuFIT\cite{nufit,Esteban:2018azc} data is drawn with red solid line. The plots show a similar nature as the correlation plots in Fig. \ref{fig:11} and \ref{fig:22}, however, the allowed regions are wider which reflects the inclusion of the experimental errors. From the first column of the figure we observe that DUNE can reject the CP conserving values ($0^\circ$, $180^\circ$ \& $360^\circ$ ) at 3$\sigma$. But when the correlation $|U_{\mu 2}|=|U_{\tau 2}|$ is considered as shown in the right column the CP conserving $\delta_{CP}$ values cannot be excluded at 3$\sigma$ . 
Note that, some of the areas allowed by the current data are disfavored by applying the correlations.
As the correlations predict a range of $\delta_{CP}$ given a set of oscillations parameters, the experiments can further constrain the range of $\delta_{CP}$ which are allowed by the present oscillation data. 
 
\subsection{Implications for Neutrino-less double \boldmath$\beta$ decay}
Neutrinoless double $\beta$ decay ($0\nu \beta \beta$) can test whether neutrinos are Majorana particles. This process takes place by emitting two electrons without the emission of the expected anti-neutrinos as observed in $2\nu \beta \beta$ decay.  
The half-life ($T_{1/2}$) for the $0\nu\beta\beta$ process is given as,
\be (T_{1/2})^{-1} = \frac{\Gamma_{0\nu\beta\beta}}{\textrm{ln}2} = G \Big|\frac{M_\nu}{m_e}\Big|^2 m_{\beta\beta}^2 , \label{T-half}\ee
where $G$ contains the lepton phase space integral, $m_e$ is the mass of electron, $M_\nu$ is the nuclear matrix element (NME) which takes into consideration all the nuclear structure effects, $m_{\beta\beta}$ stands for the effective neutrino mass and can be expressed as
\be m_{\beta\beta} = |U_{ei}^2 m_i|. \ee
Here $m_i$ are the real positive neutrino mass eigenvalues with $i=1,2,3$ for three generation and $i=1,2,3,4$ for $3+1$ neutrino mixing respectively.

Null results from several experiments have constrained the lifetimes of $0\nu\beta\beta$. KamLAND-Zen \cite{KamLAND-Zen:2016pfg} have reported a lifetime of $T_{1/2} (^{136}{Xe})  > 10.7 \times 10^{25}$ years, GERDA \cite {Agostini:2018tnm} reported as $T_{1/2} (^{76}{Ge})  > 8 \times 10^{25}$ years, CURCINO and CUORE \cite{Alduino:2017ehq} combined results reported the lifetime as $T_{1/2} (^{130}{Te}) > 1.5 \times 10^{25}$ years at $90 \%$ confidence level. The lower bound on  $T_{1/2}$ can be translated to the upper bound of effective neutrino mass ($m_{\beta\beta} $)~\cite{DellOro:2016tmg, N.:2019cot}. Using the parametrization of $U$ in Eq. \ref{maj-U4}, $m_{\beta\beta}$ can be expressed as,
\be m_{\beta\beta} \,\,=\,\, |m_1 \,c_{12}^2c_{13}^2c_{14}^2 +m_2 \, 
s_{12}^2c_{13}^2c_{14}^2e^{i2\alpha}+ m_3 \, s_{13}^2c_{14}^2 e^{i2\beta}+m_4 \,  s_{14}^2 e^{i2\gamma}|. \label{eq:mbb4n} \ee
For NH (IH) $m_1$ ($m_3$) is the lightest neutrino mass eigenstate. All other neutrino mass eigenvalues can be expressed in terms of the lightest neutrino mass and mass squared differences as follows : 
\begin{itemize}
\item  Normal Hierarchy (NH) :  $m_1 < m_2 << m_3 $ with
\begin{eqnarray}
m_2 &=& \sqrt{m_1^2\,+ \Delta m_{sol}^2} \,\,\, ; \,\,\,m_3 = \sqrt{m_1^2\, +\Delta m_{atm}^2 }\,\,\, ; \\ \nonumber
m_4 &=& \sqrt{m_1^2\,+ \Delta m_{\rm LSND}^2 },
 \label{eq:mbb4nh}
\end{eqnarray}

\item Inverted Hierarchy (IH)  :   $m_3 << m_1 \approx m_2 $ with
 \begin{eqnarray} \label{eq:mbb4ih}
m_1 &=& \sqrt{m_3^2\,+ \Delta m_{atm}^2} \,\,\, ; \,\,\,m_2 = \sqrt{m_3^2\,+ \Delta m_{sol}^2 +\Delta m_{atm}^2 } ; \\ \nonumber
m_4 &=& \sqrt{m_3^2\,+ \Delta m_{atm}^2+ \Delta m_{\rm LSND}^2 },
 \end{eqnarray}
\end{itemize}
where $\Delta m_{sol}^2 = m_2^2-m_1^2$, $\Delta m_{atm}^2 = m_3^2-m_1^2 ~(m_1^2-m_3^2) $ for NH (IH) and $\Delta m_{\rm LSND}^2 = m_4^2-m_1^2$. 

\begin{figure}[h!]
 $$
 \includegraphics[height=6.5cm,width=8cm]{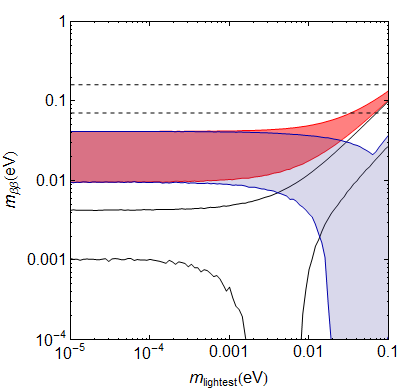}
 \includegraphics[height=6.5cm,width=8cm]{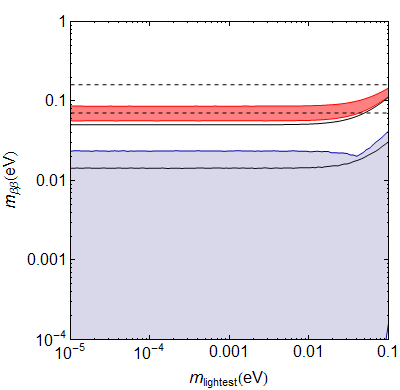}
 $$
\caption{The effective neutrino mass $m_{\beta \beta}$ for $0 \nu \beta \beta$ as a function of the lightest neutrino mass. The left panel shows the effective neutrino mass for NH while
the right panel is for IH. The red region represents  $m_{ \beta \beta} $ in presence of $\mu-\tau$ reflection symmetry in 3+1 neutrino mixing where the Majorana phases are kept as zero 
whereas the blue region represents the scenario when all the Majorana phases are fixed at $90^{\circ}$. 
The areas inside the black solid lines are the 3 neutrino allowed regions. The pair of purple dashed lines at 0.071 eV \& 0.161 eV represent the upper limit for $m_{ \beta \beta} $ by combined analysis of GERDA and KamLAND-Zen experiments.}
\label{fig:onbb}
\end{figure}

Predictions for $m_{\beta \beta}$ with respect to the lightest neutrino mass for 3+1 scheme along with the three neutrino case is presented in Fig. \ref{fig:onbb}. The plot in the left panel shows the effective neutrino mass for NH while the right panel is for IH. In generating these  plots we have varied the oscillations parameters within their 3$\sigma$ range as given in Tab. \ref{tab:data1} with $\Delta m_{\rm LSND}^2 = 1.7 ~\rm{eV^2}$~\cite{Gariazzo:2017fdh}.
In both panels of Fig. \ref{fig:onbb}, the red and blue shaded regions (corresponding to  Majorana phases fixed at $0$ and 
$90^{\circ}$ respectively) represents $m_{ \beta \beta} $ in
presence of $\mu-\tau$ reflection symmetry in 3+1 neutrino mixing.
In the plots (both left and right panel) the area between the black dashed lines at 0.071 eV \& 0.161 eV represents the upper limit for $m_{ \beta \beta} $ obtained from the 
combined analysis of GERDA and KamLAND-Zen experiments.  The width in the upper limit of $m_{\beta \beta}$ is present because of the NME uncertainty. 
For purposes of comparison we also present the three neutrino allowed regions given by the black solid lines in the Fig. \ref{fig:onbb}.
Below we discuss from  the analytic expression, the  
allowed regions for $m_{\beta \beta}$ for both hierarchies. 
The Majorana phases which are of interest to us from the point of view 
of $\mu-\tau$ reflection symmetry are 0 and $90^{\circ}$.   

\begin{itemize}
\item[*]{\bf Inverted Hierarchy:}

For inverted hierarchy, the red shaded region corresponds to the  
the case with $\alpha = \beta = \gamma=0^{\circ}$. It is seen that $m_{\beta\beta}$ 
stays almost constant in the range $0.057 - 0.087$ eV  
till $m_{\rm lightest} \sim 0.01$ eV, after which there is a slight increase in its
value.  The width of the band can be ascribed to 
the variation in the oscillation parameters in their $3\sigma$ range. 
The blue shaded region 
is obtained for $\alpha = \beta=\gamma=90^{\circ}$.  It 
is observed that complete cancellation can be obtained for these 
values of phases.  Thus the two predictions of $\mu-\tau$ reflection 
symmetry give drastically different results. 
Also, the predictions of $m_{\beta\beta}$ for the  
sterile neutrino case with 
Majorana phases as $90^{\circ}$ are   
markedly different from the three neutrino case for which there are no cancellation regions. Below we explain these features analytically in  
different limits of the lightest neutrino mass. 
 
\begin{itemize}
\item[Case 1:] When $m_3 << m_1 \approx m_2 \approx \sqrt{\Delta m_{atm}^2} $ and $m_4 \approx \sqrt{\Delta m_{\rm LSND}^2} $ we find
\begin{eqnarray}
m_{\beta\beta} = \,\, |\sqrt{\Delta m_{atm}^2 } \,c_{13}^2c_{14}^2(c_{12}^2 + s_{12}^2 e^{i2\alpha}) + \sqrt{\Delta m_{\rm LSND}^2 } \,  s_{14}^2 e^{i2\gamma} |. 
\label{eq:mbbihc1}
\end{eqnarray}
Taking the  approximations $c_{13}^2 \sim c_{14}^2 \sim 1$, $s_{12}^2 \sim 0.33 , c_{12}^2 \sim 0.67$ and $\sqrt{\Delta m_{atm}^2 } \sim 0.05~\rm{eV} , \sqrt{\Delta m_{\rm LSND}^2 } \sim 1$ eV  we obtain 
\be m_{\beta \beta} = 0.033 + 0.017 e^{i2\alpha} + s_{14}^2 e^{i2\gamma}. \label{eq:mbb4ihval}\ee
For $\alpha = \gamma =0^{\circ}$ and $s_{14}^2$ in the range 0.005-0.03 
the above gives $m_{\beta \beta}$ in the range (0.057-0.087) eV which is 
consistent with the values observed in the figure \ref{fig:onbb}.  
On the other hand for $\alpha ~ \sim {90^{\circ}}$ and $\gamma ~ \sim {90^{\circ}}$,
one can get cancellations in $m_{\beta \beta}$ for $s_{14}^2 \sim 0.016$. 
This explains the occurrence of the cancellation regions for this choice of 
phases. 

\item[Case 2:] For $m_3 \approx \sqrt{\Delta m_{atm}^2 }$,  $m_1 \approx m_2 \approx   \sqrt{2 \Delta m_{atm}^2} $ , $m_4 \approx \sqrt{\Delta m_{\rm LSND}^2} $ and we write 
\begin{eqnarray}
m_{\beta\beta} \,\,&=& \,\, | \sqrt{2 \Delta m_{atm}^2 } \,c_{13}^2c_{14}^2(c_{12}^2 + s_{12}^2 e^{i2\alpha} +  \frac{t_{13}^2}{\sqrt{2}} e^{i2\beta}) + \sqrt{\Delta m_{\rm LSND}^2 } \,  s_{14}^2 e^{i2\gamma} | . 
\label{eq:mbbihc2}
\end{eqnarray}
Again, utilizing the same values of parameters involved as in Case 1 
along with $s_{13}^2 \sim 0.024$ the effective mass can be obtained as 
\be m_{\beta \beta} = 0.047 + 0.023 e^{i2\alpha} + 0.001 e^{i2\beta} + s_{14}^2 e^{i2\gamma}.\ee

Substituting  $\alpha=\beta=\gamma=0^{\circ}$  in eq.~\ref{eq:mbbihc2} 
one gets $m_{\beta \beta}$ in the range $0.075- 0.106$ eV which can be seen 
from the figure 
for $m_{\rm lightest} \sim 0.05 $~eV.  

In this case also cancellations occur for $\alpha =  \beta = \gamma  \sim {90^{\circ}}$ and $s_{14}^2 \sim 0.023$. 
\end{itemize}

Note that in both the limits the cancellations could only be achieved because of the large value of $\sqrt{\Delta m_{\rm LSND}^2 } $. Cancellations in three generations is not possible because of absence of any term which can counter the large positive value of the first term. 
This leaves the effective neutrino mass bounded from below in the three generation case. 


It is to be noted that in the red shaded region in the right panel of Fig. \ref{fig:onbb} the value of $m_{\beta \beta}$ is $ > 0.06 \rm{eV}$, while, the current experimental bound is $m_{\beta \beta} < 0.07 \rm{eV}$.
Therefore, a portion of $m_{ \beta \beta} $ is already disfavored for certain parameter values from the current experimental bounds by GERDA and KamLAND-Zen experiments. 
\begin{figure}[h!]
 $$
 \includegraphics[height=6.5cm,width=10cm]{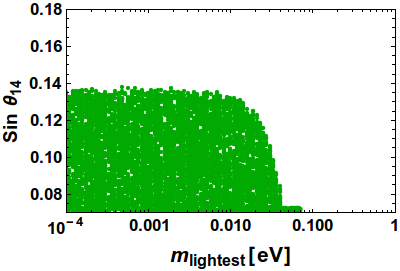}
 $$
\caption{The effective neutrino mass $m_{\beta \beta}$ for $0 \nu \beta \beta$ as a function
of $\sin \theta_{14}$.}
\label{fig:s14onbb}
\end{figure}
This experimental bound can constrain the sterile parameter $\theta_{14}$ as presented in Fig. \ref{fig:s14onbb}. In this figure the green shaded zone
represents the region allowed by GERDA and KamLAND-Zen in the $m_{\rm lightest}$ vs $\sin \theta_{14}$ plane for the case corresponding to zero Majorana phases 
of $\mu-\tau$ reflection symmetry for 3+1 neutrino mixing. We observe that $\sin \theta_{14} < 0.14$ is allowed upto $m_{\rm lightest} = 0.02~ \rm{eV}$.
As $m_{\rm lightest}$ increases further $\sin \theta_{14}$ sharply reduces because higher $m_{\rm lightest}$ is compensated by lower $\sin^2 \theta_{14}$ in order 
to satisfy the upper limit of $m_{\beta \beta}$.

It is well known that near future $0\nu\beta\beta$ experiments like  
SNO+ Phase I\cite{snophase-1}, KamLAND-Zen 800\cite{kamlamdzen-800}, and LEGEND 200\cite{Myslik:2018vts}
can test the IH region in the context of the three generation scenario. 
These experiments will be able to test the predictions for 
$\alpha=\beta=\gamma=0^{\circ}$ further. 
However, in presence of an extra sterile neutrino, a null signal 
in these experiments cannot exclude IH because of the occurence of 
the cancellation regions. 

\item[*]{\bf Normal Hierarchy:}

From the first panel of Fig. \ref{fig:onbb} we observe that 
for $\alpha=\beta=\gamma=0^{\circ}$, the red shaded region, $m_{\beta \beta}$ stays is in the range $\sim 0.01-0.05$ eV rising upto 0.1 eV for 
higher values of $m_{\rm lightest}$.  
For both  three neutrino case (region bounded by black lines)   and 3+1 neutrino mixing (with $\alpha=\beta=\gamma=90^{\circ}$, blue shaded region), complete cancellation in $m_{\beta\beta}$ is seen to occur for NH.
The cancellations occur between 0.001 eV to 0.01 eV for three neutrino mixing while it occurs between 0.01 eV to 0.1 eV for 3+1 neutrino mixing. 
Thus, compared to the three neutrino case the cancellation region for 3+1 neutrino mixing shifts towards higher values of $m_{\rm lightest}$ due to involvement of $m_4 \sim 1  \rm{eV}$. 
To analytically understand the salient features of the predictions for 
$m_{\beta \beta}$ for NH we scrutinize
the following limits:
\begin{itemize}
\item[Case 1:]
For $m_1 << \sqrt{\Delta m_{atm}^2}$ using  Eq. \ref{eq:mbb4nh} and Eq. \ref{eq:mbb4n} we obtain

\begin{eqnarray}
m_{\beta\beta} \,\,&=& \,\, |\sqrt{\Delta m_{atm}^2 } t_{13}^2 e^{i2\beta}+ \sqrt{ \Delta m_{\rm LSND}^2 } \,  s_{14}^2 e^{i2\gamma} |. 
\end{eqnarray}

In this case for $\alpha = \beta=\gamma=0^{\circ}$, the value of $m_{\beta \beta}$ 
for $t_{13}^2 \sim 0.024$, $s_{14}^2 \sim 0.05$, $\sqrt{\Delta m_{atm}^2 } \sim 0.05$ eV we get $m_{\beta \beta} \sim 0.03$ eV which is in the range 
obtained in the figure \ref{fig:onbb}. 
Since  
$t_{13}^2 \sim s_{14}^2$ and $\Delta m_{atm}^2 << \Delta m_{\rm LSND}^2$ cancellations are not possible for smaller values of  $m_1$. 
For $\alpha=\beta=\gamma =90^{\circ}$ gives similar results as the $\alpha = \beta=\gamma=0^{\circ}$ case which is also corroborated from the figure.

\item[Case 2:]
Now, when $m_1 \sim \sqrt{\Delta m_{atm}^2} $ then the expression for the effective mass reduces to 
\begin{eqnarray}
m_{\beta\beta} \,\,&=& \,\, |\,c_{13}^2c_{14}^2 \sqrt{ \Delta m_{atm}^2 }(c_{12}^2 +   s_{12}^2 e^{i2\alpha} + \sqrt{2} t_{13}^2 e^{i2\beta}) \\ \nonumber
& & + \sqrt{\Delta m_{\rm LSND}^2 } \,  s_{14}^2 e^{i2\gamma} |, \\ 
&\approx & 0.033 + 0.017 e^{i2\alpha} + 0.002 e^{i2\beta} + s_{14}^2 e^{i2\gamma}.
\end{eqnarray}

For zero values of the Majorana phases $m_{\beta \beta}$ can reach upto 0.08 eV 
for $s_{14}^2 = 0.03$.  
In this limit, cancellations can occur for 
$\alpha = \beta = \gamma = 90^{\circ}$ and $s_{14}^2 \sim 0.015$. 
\end{itemize} 

Note that the predictions for $m_{\beta \beta}$ in the 3+1 picture for zero Majorana phases for higher values of $m_{\rm lightest}$ are already crossing the current experimental values and this can 
put bound on sterile parameters as in the previous case. 
Since the values of $m_{\beta \beta}$ upto $m_{\rm lightest} \sim 0.001$ eV 
is in the same ballpark as the IH values for  three generation scenario, 
the near future experiments designed to test the 3 generation IH region 
can also probe this region. For representative purposes we have included the predictions for $m_{\beta\beta}$ for various cases for few benchmark points of the
lightest neutrino mass ($m_1$ for NH and $m_3$ for IH) in Table \ref{nobbtab}.

\begin{table}[]
\centering
\resizebox{17cm}{!}{%
\begin{tabular}{|l|l|c|c|c|c|c|c|}
\hline
\multicolumn{2}{|c|}{\multirow{2}{*}{}} & \multicolumn{3}{c|}{$m_{\beta\beta}$ for NH} & \multicolumn{3}{c|}{$m_{\beta\beta}$ for IH} \\ \cline{3-8} 
\multicolumn{2}{|c|}{}                                   &   $m_1=0.001$ eV                      &  $m_1=0.01$ eV            &    $m_1=0.052$ eV   &   $m_3=0.001$ eV    &   $m_3=0.01$ eV    &   $m_3=0.052$ eV     \\ \hline
\multirow{2}{*}{}    Sterile : Majorana Phases = 0       &  ${m_{\beta\beta}}_{\rm min} \times 10^{-3}$  &  10.28  & 17.01 & 57.70 & 56.43 & 57.19 & 77.28 \\ \cline{2-8} 
                                                         &  ${m_{\beta\beta}}_{\rm max} \times 10^{-3}$  &  42.15  & 48.40 & 89.27 & 86.47 & 87.37 & 108.58 \\ \hline
\multirow{2}{*}{}   Sterile : Majorana Phases = $90^{\circ}$  &  ${m_{\beta\beta}}_{\rm min} \times 10^{-3}$  &  8.92   & 3.82  & 0.03  & 0.01  & 0.02  & 0.01\\ \cline{2-8} 
                                                         &  ${m_{\beta\beta}}_{\rm max} \times 10^{-3}$  &  40.94  & 36.76 & 20.40 & 23.68 & 23.15 & 24.54  \\ \hline
\multirow{2}{*}{}    3 Generation                        &  ${m_{\beta\beta}}_{\rm min} \times 10^{-3}$  &  0.46   & 0.76  & 13.77 & 14.28 & 14.50 & 20.23 \\ \cline{2-8} 
                                                         &  ${m_{\beta\beta}}_{\rm max} \times 10^{-3}$  &  4.83   & 12.10 & 53.18 & 50.37 & 51.40 & 73.02 \\ \hline
\end{tabular}
}\label{nobbtab}
\caption{Predictions for $m_{\beta\beta}$ for various cases for few benchmark points of the lightest neutrino mass ($m_1$ for NH and $m_3$ for IH). ${m_{\beta\beta}}_{\rm min}$ and 
${m_{\beta\beta}}_{\rm min}$ stands for the respective minimum and maximum values of $m_{\beta\beta}$ for each benchmark point.}
\end{table}


In our analysis we have considered $\Delta m^2_{LSND} = 1.7~ {\rm eV^2}$ which gives the physical mass of sterile neutrino $m_4 = m_{ph}^s \sim 1.3~ {\rm eV}$.
The Cosmic Microwave Background analysis in $\Lambda_{\rm CDM} + r_{0.05} + N_{\rm eff} + m_{\rm eff}^s$ model using the Planck 2015 data~\cite{Ade:2015xua} gives the $N_{\rm eff} < 3.78$ and $m_{\rm eff}^s < 0.78~\rm{eV}$~\cite{Choudhury:2018sbz}. The bounds including other datasets are more stringent than this.
As the effective mass in terms of the  $N_{\rm eff}$ and physical mass of sterile neutrino is given as $m_{\rm eff}^s = \Delta N_{\rm eff}^{3/4}~ m_{\rm ph}^s$, where, $\Delta N_{\rm eff} = N_{\rm eff} - 3.046$  one gets $m_{\rm ph}^s < 0.98$ eV at $95\%$ CL. 

\end{itemize}

\section{Conclusion}\label{sec5}
To understand the observed pattern of lepton mixing, $\mu-\tau$ symmetry may play a crucial role as
it can be originated from various discrete flavor symmetries. Along with three active neutrinos, 
presence of one sterile neutrino may have some interesting predictions on neutrino mixing  angles 
and Dirac CP violating phases within the framework of such $\mu-\tau$ symmetries. Conventional 
$\mu-\tau$ permutation symmetry for 3+1 picture is not a phenomenologically viable scenario as it
can not  explain correct neutrino mixing (since it predicts $\theta_{13}=0^\circ$).  Hence here we have analyzed a simple extension of it, known as $\mu-\tau$ reflection symmetry, in the context of 3+1 neutrino mixing.
We formulate the mass matrix compatible with the lepton mixing matrix which can give rise to $\mu-\tau$ 
reflection symmetry, defined via $|U_{\mu i}|=|U_{\tau i}|~{\rm where}~{i=1,2,3,4}$. We obtain and plot
the correlations connecting the mixing angle $\theta_{23}$ and the CP phase $\delta$ for the case when
sterile phases are assumed to be zero, as well as  present the correlation plots with the sterile phases
varied in their full range.
We find that if we consider total $\mu-\tau$ reflection symmetry i.e. $|U_{\mu i}|=|U_{\tau i}|$ is
simultaneously satisfied for all the four columns then the mixing angle $\theta_{23}$ is confined in a
 narrow region around $\theta_{23} = 45^\circ$ and $\delta$ is restricted around the maximal CP violating 
 values. However, the deviation of $\theta_{23}$ from maximal value with the inclusion of the sterile mixing
 is not sufficient to account for the observed best fit value. This prompts us to consider partial $\mu-\tau$ 
 reflection symmetry and study the consequences for each column individually. 
 
The equalities $|U_{\mu 1}|=|U_{\tau 1}|$ and $|U_{\mu 2}|=|U_{\tau 2}|$ yield important correlations among the
neutrino mixing angle $\theta_{23}$ and Dirac CP phase $\delta$.  Interestingly we find that the best-fit value 
for ($\theta_{23},\delta$)  shows a good agreement with inverted neutrino mass hierarchy for $|U_{\mu 1}|=|U_{\tau 1}|$
and normal mass hierarchy for  $|U_{\mu 2}|=|U_{\tau 2}|$. With precise measurement of $\theta_{23}$ and $ \delta$ 
in near future there is a clear possibility of verifying these correlations. The inclusion of the sterile CP phases
widens the allowed regions. For the equality $|U_{\mu 3}|=|U_{\tau 3}|$ the mixing angle $\theta_{23}$ sightly deviates
from its maximal value and falls mostly in the lower octant ($\theta_{23}<45^{\circ}$) including the effect of sterile
mixing angle. This, however, is not supported by the global oscillation analysis~\cite{nufit,Esteban:2018azc,Capozzi:2018ubv}.
The equality $|U_{\mu 4}|=|U_{\tau 4}|$ yields an one-to-one correspondence between the sterile mixing angles $\theta_{24}$
and $\theta_{34}$ making it one of the most significant finding in the present study. So far, there exists only an upper
limit on $\theta_{34}$. Interestingly, here we find that the correlation obtained from $|U_{\mu 4}|=|U_{\tau 4}|$ 
restricts $\theta_{34}$ within the range $4.9^{\circ}-9.8^{\circ}$.
The allowed region in the $\theta_{24}-\theta_{34}$ plane also gets severely 
restricted. Future experiments sensitive to these mixing angles can test the correlations discussed.

We also explore the possibility of testing the $\mu$-$\tau$ reflection symmetry for 3+1 neutrino mixing at the future LBL experiment DUNE. 
The application of the correlations constrains a significant area  of the parameter space yet unconstrained by the present global fit data. In particular the constraint is more stringent for the  relation $|U_{\mu 1}|=|U_{\tau 1}|$ and all the CP conserving values $\delta ~ = 0^\circ, 180^\circ, 360^\circ$ are excluded at $3\sigma$. However, for $|U_{\mu 2}|=|U_{\tau 2}|$ CP conserving values of $\delta$ remain allowed.

Furthermore, we expound the implications of the $\mu$-$\tau$ reflection symmetry for 3+1 neutrino mixing at neutrinoless double $\beta$ experiments by calculating the effective neutrino mass ($m_{\beta \beta}$) for this scenario. 
The $\mu$-$\tau$ reflection symmetry predicts the Majorana phases to be zero
or $90^{\circ}$. For the former predicted effective neutrino mass is higher and can be explored at the future $0 \nu \beta \beta$ experiments. 
In fact for higher values of the lightest neutrino mass
the effective neutrino mass $m_{\beta \beta}$ can cross
the current experimental bounds when Majorana phases are assumed to be zero. 
We have shown that, 
for IH, this  constrains the sterile parameter $\theta_{14}$ under $8^\circ$ using the bounds on the effective neutrino mass.
For the Majorana phases as $90^{\circ}$, for IH there can be cancellation regions 
in stark contrast with the three generation predictions. For NH, the 
cancellation region for the 3+1 case occur for higher values of the lightest 
neutrino mass as compared to the three neutrino picture. 

In conclusion, $\mu-\tau$ reflection symmetry for sterile neutrinos 
in a 3+1 picture gives some interesting predictions 
which can be tested in future neutrino oscillation and neutrinoless double 
beta experiments and the scenario can be confirmed or falsified.  


\acknowledgments
The authors thank Anjan S. Joshipura, Ketan M. Patel and Tanmay Poddar 
for useful discussions and suggestions. The authors also thank Vishnudath K.N. 
for his help with the numerical work.


\bibliography{mutau-flav-ref.bib}

\end{document}